\documentclass[prb,aps,amsmath,amssymb,showpacs,twocolumn]{revtex4-1}
\usepackage{amsmath,amssymb,graphicx,float,dcolumn,bm,color,colortbl,multirow}
\usepackage[titletoc,title]{appendix}
\usepackage[dotinlabels]{titletoc}
\usepackage{cancel}
\usepackage{braket}
\usepackage{mathrsfs}
\usepackage[mathscr]{euscript}
\usepackage{hyperref}
\usepackage{comment}
\usepackage[dvipsnames]{xcolor}

\newcolumntype{P}[1]{>{\centering\arraybackslash}p{#1}}
\newcommand{\bsym}[1]{\boldsymbol{#1}}
\newcommand{\rbkt}[1]{\left( #1\right)}
\newcommand{\sbkt}[1]{\left[ #1\right]}

\newcommand{\tdots}{.\,.\,}

\newcommand{\iprod}[2]{\left\langle #1 \right| \left. #2 \right\rangle}
\newcommand{\proj}{\mathcal{P}}
\newcommand{\secref}[1]{~\ref{#1}}

\begin{document}

\title{Real-space entanglement spectra of projected fractional quantum Hall states using Monte Carlo methods}
\author{Abhishek Anand and G. J. Sreejith}
\affiliation{Indian Institute of Science Education and Research, Pune 411008, India}

\begin{abstract} 
Real-space entanglement spectrum (RSES)  of a quantum Hall (QH) wavefunction gives a natural route to infer the nature of its edge excitations. Computation of RSES becomes expensive with an increase in the number of particles and included Landau levels (LL). RSES can be efficiently computed using Monte Carlo (MC) methods for trial states that can be written as products of determinants such as the composite fermion (CF) and parton states. This computational efficiency also applies to the RSES of lowest Landau level (LLL) projected CF and parton states; however, LLL projection to be used here requires approximations that generalize the Jain Kamilla (JK) projection.
This work is a careful study of how this approximation should be made. We identify the approximation closest in spirit to JK projection and perform tests of the approximations involved in the projection by comparing the MC results with the RSES obtained from computationally expensive but exact methods. We present the techniques and use them to calculate the exact RSES of the exact LLL projected bosonic Jain $2/3$ state in bipartition of systems of sizes up to $N=24$ on the sphere. For the lowest few angular momentum sectors of the RSES, we present evidence to show that MC results closely match the exact spectra. We also discuss other plausible projection schemes. We also calculate the exact RSES of the unprojected fermionic Jain $2/5$ state obtained from the exact diagonalization of the Trugman-Kivelson Hamiltonian in the two lowest LLs on the sphere. By comparing with the RSES of the unprojected $2/5$ state from Monte Carlo methods, we show that the latter is practically exact.
\end{abstract}
\maketitle

\section{Introduction}
Fractional quantum Hall (FQH) effect\cite{Klitzing80,Tsui82} provides a rich set of experimentally realizable interacting topologically ordered quantum phases. Conceptual insights into the microscopic structure of these phases have been aided by careful numerical studies of variational wavefunctions describing the phases.
The many aspects of the FQH states like fractional charge,\cite{Laughlin83,Arovas84} non-trivial statistics\cite{Leinaas77,Arovas84,Kjonsberg99b,Jeon03b,Tserkovnyak03} and edge modes have been understood in the language of such variational wavefunctions. Entanglement spectrum (ES)\cite{Li08} of a variational state is a key tool in the characterization of the order in the state.\cite{Calabrese08, Thomale10, Cirac11, Pollmann10,Regnault2013} The notion of entanglement spectra was introduced as a means to get more insights into the topological order beyond what is provided by more succinct measures like the entanglement entropy.\cite{Kitaev06a,Haque07a,Haque07b,SierraRodriguez2009} Entanglement spectra of a many-body state is defined as the eigenvalue spectrum (typically represented in a negative log scale) of the reduced density matrix after a bipartition of the system. Correspondence between the entanglement Hamiltonian and the edge spectrum has been explored both in the FQH contexts and elsewhere. \cite{Fidkowski10,turner2010,yao2010,ludwig12,Pollmann10,Poilblanc10,Peschel2011} Being a quantity naturally obtained during DMRG calculations, entanglement spectrum has been used to characterize the phases of interacting Hamiltonians in such studies.\cite{Sheng11b,Zaletel12,Zaletel13,Sheng15}

In quantum Hall systems, entanglement spectrum can be defined by bipartitioning the system in the momentum  orbital space or in the real space producing the orbital space entanglement spectrum (OES) or real space entanglement spectrum(RSES) respectively. Since under Landau quantization, the single particle momentum orbitals are spatially localized, they contain closely related information. Both OES and RSES have been studied extensively for many QH states.\cite{Chandran2012,Sterdyniak11,Rodriguez12,Greg2021,Rodriguez13,Anand22} Typically OES is easier to compute for states obtained from exact diagonalization or DMRG calculations whereas RSES is easier to compute for states described by variational wavefunctions expressed in terms of particle coordinates.
In this paper, we focus on the RSES estimations using Monte Carlo methods focusing on the Jain composite fermion (CF) states.\cite{Jain89}

ES of CF wavefunctions have been studied  but methods using exact evaluation of the LLL projected state is computationally expensive, allowing studies only in small systems ($N\sim 10$).\cite{Sterdyniak11} On the other hand, entanglement spectrum of the unprojected CF state and parton states\cite{Jain89} can be efficiently obtained using Monte Carlo methods.\cite{Rodriguez13,Anand22}
The bottleneck in extending the same technique to the projected CF states (which are energetically more favorable) is the difficulty in implementing the LLL projection.
In a large number of studies involving energetics of the CF states, an approximate method introduced by Jain and Kamilla (JK) to perform lowest Landau level (LLL) projection has been found to provide computationally efficient and reliable results.\cite{Jain97,Jain97b,Moller05}
Combining the JK projection with the MC methods for RSES evaluation involves further approximations but this can be used for calculations in systems upto hundred particles.\cite{Rodriguez13,Greg2021,Anand22}
Testing the approximations require comparison with computationally expensive but exact RSES calculated using alternate methods. This is the goal of the paper.

We identify cases where alternate exact methods can be employed to calculate RSES in relatively large systems.
For the fermionic Jain $2/5$th state, the Trugman-Kivelson (TK) Hamiltonian\cite{Trugman85} can be diagonalized to produce an exact expansion of the unprojected CF state. RSES of the state can be obtained using a generalization of the method presented by Chandran $\textit{et al.}$ in Ref.~\onlinecite{Chandran2012} We use this to show that the MC method indeed produces practically exact results for the RSES.

Construction of exact Hamiltonians for the projected state is an open problem,\cite{Sreejith18} so a similar strategy cannot be employed to test the RSES for the projected state. Instead we consider the case of the bosonic Jain $2/3$rd state, where we employ exact projection in a manner that allows us to exactly compute the low momentum entanglement spectrum in systems as large as $N=24$. We find that as system size increases the results from approximate projection employed in the MC method approach the exact RSES, at least in the low momentum sectors.

We emphasize that the results presented in this work is not a comparison of the RSES of the exact projected CF states ($\psi_{\rm EX}$) and the RSES of the JK projected CF states ($\psi_{\rm JK}$). Since the two states are nearly identically to each other, we expect their RSES to be nearly the same. Instead what we are testing is the effect of the approximate projection used while implementing the MC method for RSES calculation. The approximation in the latter is similar in spirit to the JK projection but is not the same.
Secondly, the results presented compare the approximate results from MC estimates of RSES with the RSES of the $\psi_{\rm EX}$. The comparison is done with $\psi_{\rm EX}$ rather than with $\psi_{\rm JK}$ because, using methods described in Sec. \ref{exactProj}, RSES of $\psi_{\rm EX}$ can be computed exactly at least for small systems. Doing the same for $\psi_{\rm JK}$ is harder.

This paper is structured as follows. We begin by presenting the numerical techniques involved.
We present, in Sec~\ref{RSESusingED}, a strategy for a numerically exact computation of the RSES of the unprojected Jain $2/5$th state obtained from exact diagonalization of the  TK Hamiltonian. We will later use this to demonstrate that the RSES using MC method is practically exact for the unprojected state.
In   Sec.\secref{RSESAlgo}, we give a summary of the method for computing RSES of variational states by expanding them in terms of entanglement wavefunctions (EWFs). The details of the method, originally introduced in Ref.~\onlinecite{Rodriguez13}, can be found in Refs.~\onlinecite{Greg2021,Anand22}.
Section \secref{exactProj} provides the details for using this method with exactly projected Jain $2/3$rd state of bosons. We could use this to obtain numerically exact RSES in systems up to size $24$.
Section \secref{approxProj} details the approximations that are made to perform LLL projection of the EWFs, which makes accessing large systems possible. 
All numerical results benchmarking the methods are given in Sec.\secref{NumRes}, and finally we conclude with Sec.\secref{conclusion}.

{\it Notations:} All calculations in the paper are performed for systems in the  spherical geometry where the single particle Landau orbitals for a particle in the $n$th LL and with angular momentum $m$ are the monopole harmonics,\cite{Wu76,Wu77,Haldane83} given by
\begin{align} \label{yqlm}
Y_{Qnm}=&N_{Qnm}(-1)^{Q+n-m}v^{Q-m}u^{Q+m} \nonumber \\
 & \times \sum_{s=0}^{n} {n \choose s} {2Q+n \choose Q+n-m-s}|v|^{2(n-s)}|u|^{2s}
\end{align}
where $u$ and $v$ are given by $u=\text{cos}(\theta/2)e^{\iota\phi/2}$ and $v=\text{sin}(\theta/2)e^{-\iota\phi/2}$ in terms of coordinates $0\leq\theta<\pi$ and $0\leq\phi<2\pi$ on the sphere, and $Q$ quantifies the strength of monopole which produces a radial magnetic field of flux $2Q$ in units of flux quanta $\phi_0=hc/e$. The normalization factor is given by 
\begin{multline}
N_{Qnm}= \\
\rbkt{\frac{(2Q+2n+1)}{4\pi}\frac{(Q+n-m)!(Q+n+m)!}{n!(2Q+n)!}}^{1/2}
\end{multline}

\section{RSES of state expanded in Slater determinant basis} \label{RSESusingED}
In this section, we describe the method to calculate RSES for fermionic quantum Hall states which are expressed as linear combinations
\begin{equation} \label{EDstate}
\psi(\bsym{r}_1,\dots,\bsym{r}_N)=\sum_{\bsym{\lambda}}\,c_{\bsym{\lambda}} \mathcal{M}_{\bsym{\lambda}} (\bsym{r}_1,\dots,\bsym{r}_N)
\end{equation}
where the basis states $\mathcal{M}_{\bsym{\lambda}}$ are Slater determinants of single particle momentum orbitals.
The coefficients $c_{\bsym{\lambda}}$'s can, for instance, be from exact diagonalization.
The basis states $\mathcal{M}_{\bsym{\lambda}}$ are parametrized by the ordered list of occupied single particle orbitals $\bsym{\lambda}\equiv (\lambda_1,\lambda_2\dots)$ and can be expanded as
\begin{align}\label{SlaterState}
\mathcal{M}_{\bsym{\lambda}}(\bsym{r}_1,\dots,\bsym{r}_{N})=\frac{1}{\sqrt{N!}}\, \sum_{\sigma \in \mathcal{S}_N} \epsilon (\sigma) \prod_{i=1}^{N}\phi_{\lambda_{\sigma(i)}}(\bsym{r}_i) 
\end{align}
where $\phi_{\lambda_i}(\bsym{r}_i)$'s are the normalized single particle Landau orbitals $\lambda_i\equiv(n_i,m_i)$ specified by the LL-index $n_i$ and the angular momentum $m_i$, and $\mathcal{S}_N$ is the set of all permutations of $(1,2,\dots,N)$. In the spherical geometry, the single particle orbitals are given by monopole harmonics given in Eq.\eqref{yqlm}.

We present a method which enables us to compute RSES for states with particles occupying different LLs. For a real-space cut which respects the rotational symmetry of the system, angular momentum states (projected onto either subsystem) remain orthogonal to each other as long as they are in same LL. However states with same angular momentum but different LLs have non-zero overlap due to restricted limits of integration within each subsystem. The method presented below extends the one given in Ref. \onlinecite{Chandran2012} which works for states restricted to the LLL, by incorporating non-orthogonal momentum states.

In this work, we will use this method to compute the RSES of unprojected Jain $2/5$-state, which we get as the ground state (using ED) of Trugman-Kivelson Hamiltonian\cite{Trugman85} projected into the lowest $2$ LLs treated as degenerate.

\begin{figure}[h!]
	\includegraphics[width=0.4\columnwidth]{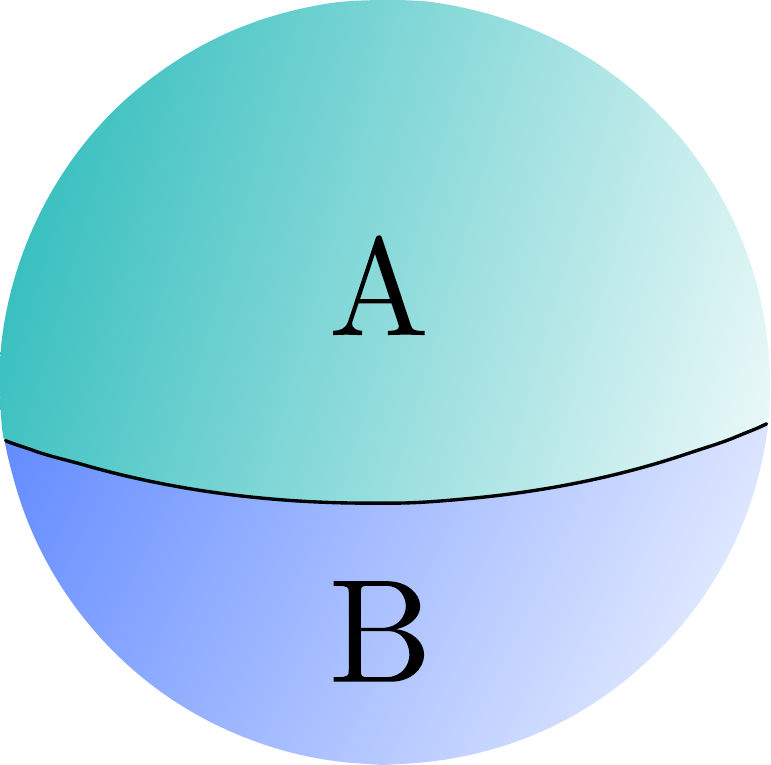}
	\caption{ Azimuthally symmetric cut for the spherical geometry. \label{fig:schematic}}
\end{figure}

For any wavefunction $\psi(\bsym{r}_1,\dots,\bsym{r}_N)$ for $N$ particles, the density matrix is given by
\begin{equation} \label{eqRho}
    \rho(\bsym{r}'_1,\tdots,\bsym{r}'_N;\bsym{r}_1,\tdots,\bsym{r}_N) = \frac{\bar{\psi}(\bsym{r}_1,\tdots,\bsym{r}_N)\psi(\bsym{r}'_1,\tdots,\bsym{r}'_N)}{\int\prod_{i} \text{d}^2\bsym{r}_i\,|\psi(\bsym{r}_1,\tdots,\bsym{r}_N)|^2}
\end{equation}
We partition the system  into two subsystems $A$ and $B$ using an azimuthally symmetric cut (Fig.\ref{fig:schematic}) and consider the sector where region $A$ contains  $N_A$ particles (and $B$ has $N_B=N-N_A$ particles). We use the following shorthands for collections of particle coordinates $\bsym{R}\equiv (\bsym{r}_1,\dots,\bsym{r}_N)$, $\bsym{R}_A\equiv (\bsym{r}_1,\dots,\bsym{r}_{N_A})$ and $\bsym{R}_B\equiv (\bsym{r}_{N_A+1},\dots,\bsym{r}_{N})$.
The reduced density matrix for subsystem $A$ is then given by

 \begin{align} \label{eqRhoA}
    \rho_{N_A}(\bsym{R}^{'}_{A};\bsym{R}_{A}) =& \frac{ \int_B \text{d}\bsym{R}_B\,  \rho(\bsym{R}_{A},\bsym{R}_{B}; \bsym{R}^{'}_{A},\bsym{R}_{B})}{\int_A \text{d}\bsym{R}_A\int_B \text{d}\bsym{R}_B\, \rho(\bsym{R}_{A},\bsym{R}_{B};\bsym{R}_{A},\bsym{R}_{B}) } \nonumber \\
    =& \frac{1}{p_{N_A}} {N \choose N_A} \int_B \text{d}\bsym{R}_B\,  \rho(\bsym{R}_A,\bsym{R}_B; \bsym{R}^{'}_A,\bsym{R}_B)
\end{align}
where $p_{N_A}$ is the probability that subsystem $A$ contains exactly $N_A$ particles. Using Eqs.\eqref{EDstate} and \eqref{SlaterState}, we can rewrite the Eq.\eqref{eqRhoA} as 
 \begin{align} \label{eqRhoA2}
    \rho_{N_A} =& \frac{1}{p_{N_A}} {N \choose N_A}  \sum_{\bsym{\lambda},\bsym{\lambda}'}\,  \bar{c}_{\bsym{\lambda}}  c_{\bsym{\lambda}'} \int_B \text{d}\bsym{R}_B\,  \overline{\mathcal{M}}_{\bsym{\lambda}} \mathcal{M}_{\bsym{\lambda}'}
\end{align}
To simplify further, we will use the following property

\begin{align}   \label{SlaterProd}
\mathcal{M}_{\bsym{\lambda}}(\bsym{R}) = \sqrt{ \frac{N_A!N_B!}{N!}}\, \sum_{\substack{\bsym{\mu},\bsym{\nu}\\ \langle\bsym\mu;\bsym\nu \rangle=\bsym{\lambda} } }\, \epsilon_{\bsym{\mu}\bsym{\nu}} \,\mathcal{M}_{\bsym{\mu}}(\bsym{R}_A )\mathcal{M}_{\bsym{\nu}}(\bsym{R}_{B}).
\end{align}
Here the Slater determinant of $N$ particles is expanded as anti-symmetrization of products of Slater determinants corresponding to ordered set of orbitals $\bsym\mu$ (of size $N_A$) and $\bsym\nu$ (of size $N_B$) such that the ordered combination of two is equal to $\bsym\lambda$ ( this constraint is represented by  $\langle\bsym\mu;\bsym\nu \rangle=\bsym{\lambda}$). The sign corresponding to the permutation $\sigma$, which makes   $(\lambda_{\sigma(1)},\dots,\lambda_{\sigma(N)})=(\mu_1,\dots,\mu_{N_A},\nu_{N_A+1},\dots,\nu_{N})$ is given by $\epsilon_{\bsym{\mu}\bsym{\nu}}=\epsilon(\sigma)$. This allows us to rewrite $\rho_{N_A}$ as

\begin{align} \label{eqRhoA3}
\rho_{N_A}(\bsym{R}^{'}_A,\bsym{R}_A) =&  \sum_{\bsym\mu, \bsym\mu'}\,   Q_{\bsym{\mu},\bsym{\mu}'}\overline{\mathcal{M}}_{\bsym{\mu}}(\bsym{R}_A) 	 \mathcal{M}_{\bsym{\mu}'}(\bsym{R}^{'}_A)
\end{align}
where
\begin{gather} \label{Qmat}
Q_{\bsym{\mu},\bsym{\mu}'}=\, \frac{1}{p_{N_A}}\sum_{ \bsym\nu,\bsym\nu'}\  \int_B\text{d}\bsym{R}_B\, {\mathcal{\bar F}^{\bsym\mu}}_{\bsym{\nu}}(\bsym{R}_B) \mathcal{F}^{\bsym\mu'}_{\bsym{\nu}'}(\bsym{R}_B) \nonumber \\
\mathcal{F}^{\bsym\mu}_{\bsym{\nu}}(\bsym{R}_B)=\, \epsilon_{\bsym{\mu}\bsym{\nu}} c_{\langle\bsym\mu;\bsym\nu \rangle}\mathcal{M}_{\bsym\nu}(\bsym{R}_B)  \label{Qmatrix}
\end{gather}
 This integral contains the overlaps between the angular momentum states which are restricted in $B$ subsystem. As mentioned before, overlap between single particle momentum orbitals in subsystem $B$ is non-zero only when they have same angular momentum. Hence, only those ordered sets $\bsym{\nu},\bsym{\nu}'$ contribute in the sum in Eq.\eqref{Qmat} which have identical set of angular momentum quantum numbers. Note that the LL-indices need not be the same. The matrix $Q$ in Eq.\eqref{Qmatrix} can be numerically computed by summing over such ordered sets $\bsym{\nu}$ and $\bsym{\nu}'$.

The reduced density matrix $\rho_{N_A}$ is block-diagonal in angular momentum. Its $L_z^A$-sectors, represented as $\rho_{N_A,L_z^A}$ is 
\begin{align} \label{eqRhoA4}
\rho_{N_A,L_z^A} (\bsym{R}^{'}_A,\bsym{R}_A) =& \sum^{'}_{ \bsym\mu, \bsym\mu'}\,  \overline{\mathcal{M}}_{\bsym{\mu}} (\bsym{R}_A) 	Q_{\bsym{\mu},\bsym{\mu}'} \mathcal{M}_{\bsym{\mu}'}(\bsym{R}^{'}_A).
\end{align}
where we the restricted sum is over ordered sets $\bsym{\mu},\ \bsym{\mu}'$ of size $N_A$ which have the correct total angular momentum {\it i.e.}  $L_z(\bsym{\mu})=L_z(\bsym{\mu}')=L_z^A$. As the Slater determinant states $\mathcal{M}_{\bsym{\mu}}$s span the entire Hilbert space of $A$-subsystem, any eigenvector $\chi$  of $\rho_{N_A}$ with eigenvalue $k$ can be written as following linear combination
\begin{equation} \label{eqRhoA5}
	\chi(\bsym{R}_A) =\sum_{\bsym{\mu}} a_{\bsym{\mu}}\mathcal{M}_{\bsym{\mu}}(\bsym{R}_A)
\end{equation}
where the basis $\mathcal{M}_{\bsym{\mu}}$s are generally not orthogonal if they contain states from higher LLs $(n>1)$. Using the eigenvalue equation, given by
\begin{align} 
\int_A\text{d}\bsym{R}_A\, \rho_{N_A}(\bsym{R}_A,\bsym{R}^{'}_A)\chi(\bsym{R}_A)   =& k \chi(\bsym{R}^{'}_A)
\end{align}
and Eq.\eqref{eqRhoA5}, it can be shown that a matrix $M$ can be constructed such that it has same set of non-zero eigenvalues as that of $\rho_{N_A}$, where $M = QP$ where $Q$ is defined in Eq.\eqref{Qmatrix} and $P$ is given by
\begin{equation} 
P_{\bsym{\mu},\bsym{\mu}'}= \int_A\text{d}\bsym{R}_A\, \overline{\mathcal{M}}_{\bsym{\mu}}(\bsym{R}_A) \mathcal{M}_{\bsym{\mu}'}(\bsym{R}_A) 
\end{equation}
which is the overlap matrix (inside $A$) for EWFs, which can be computed using accurate numerical integration.

The method allows exact computation of RSES for ED eigenstates of various parent QH Hamiltonians where particles are allowed to occupy different LLs.\cite{Haldane83,Trugman85,Bandyopadhyay18,Anand21} Exponential growth of dimension restricts the usage of this method to only smaller systems $(N \lesssim 10)$. Nonetheless, having exact RSES at disposal allows us to benchmark the MC method. In the next section, we  discuss an efficient RSES computation method based on the Monte Carlo technique. 


\section{RSES using Entanglement Wavefunctions} \label{RSESAlgo}
Many FQH phases are described by wavefucntions which have a product-of-Slater-determinants form, for instance, the Jain CF states\cite{Jain89} and the partonic QH states.\cite{Jain89b} An efficient algorithm for RSES computation has already been studied for such QH states, using Monte Carlo method in Ref.~\onlinecite{Rodriguez13}. In this section we present a brief overview of the method for unprojected CF states first followed by description of strategies to deal with projected states.

\subsection{RSES of unprojected CF states}\label{RSESunproj}

The unprojected CF state for $N$-particles at filling fraction $\nu=n/(np+1)$ has the form
\begin{equation} \label{unprojCF}
	\psi^{\rm unproj}_{n/(np+1)}(\bsym{R})= \sbkt{ \Phi_1(\bsym{R})}^{p}\Phi_n(\bsym{R})
\end{equation}
where $\Phi_n$ is a Slater determinant state where $N$ particles completely fill $n$ LLs. As before, we assume a rotationally symmetric cut in real space which restricts $N_A$ particles in region $A$ and $N_B=N-N_A$ in the $B$ region. Using Eq.\eqref{SlaterProd}, one can rewrite the state as
\begin{equation}
\psi^{\rm unproj}_{n/(np+1)}(\bsym{R})= {\rm anti}\sum_{\vec{\bsym\mu},\vec{\bsym\nu}}s\rbkt{\vec{\bsym{\mu}};\vec{\bsym{\nu}}}
\xi_{\vec{\bsym\mu}}^{A}(\bsym{R}_A) \xi_{\vec{\bsym\nu}}^{B}(\bsym{R}_B)
\end{equation}
where $\vec{\bsym{\mu}}$ and $\vec{\bsym{\nu}}$ are collections of $p+1$ ordered sets of orbitals $\vec{\bsym{\mu}}=\rbkt{\bsym{\mu}_1,\dots,\bsym{\mu}_{p},\bsym{\mu}^*}$ and $\vec{\bsym{\nu}}=\rbkt{\bsym{\nu}_1,\dots,\bsym{\nu}_{p},\bsym{\nu}^*}$. Here 
$\bsym\mu_i$ and $\bsym\nu_i$ (for each $i=1,\dots,p$) are disjoint ordered sets of orbitals, of length $N_A$ and $N_B$ respectively, containing single particle orbitals from the Slater determinant $\Phi_1$.
Similarly, $\bsym\mu^*$ and $\bsym\nu^*$ are disjoint ordered sets of lengths $N_A$ and $N_B$ made of occupied single-particle orbitals from $\Phi_n$.
Integer valued combinatorical factors $s\rbkt{\vec{\bsym{\mu}};\vec{\bsym{\nu}}}$ arise from considering the signs in Eq. \ref{SlaterProd}. The entanglement wavefunctions (EWFs) $\xi_{\vec{\bsym\mu}}^{A}$ and $\xi_{\vec{\bsym\nu}}^{B}$ are defined as
\begin{align} \label{EWFexpansion}
\xi_{\vec{\bsym\mu}}^{A}(\bsym{R}_A)=& \mathcal{M}^{A}_{\bsym\mu^*}(\bsym{R}_A)\prod_{i=1}^p \mathcal{M}^{A}_{\bsym\mu_i} (\bsym{R}_A)  \nonumber \\
\xi_{\vec{\bsym\nu}}^{B}(\bsym{R}_B)=& \mathcal{M}^{B}_{\bsym\nu^*}(\bsym{R}_B)\prod_{i=1}^p \mathcal{M}^{B}_{\bsym\nu_i}(\bsym{R}_B) 
\end{align}
where $\mathcal{M}^{A}_{\bsym\mu}$ and $\mathcal{M}^{B}_{\bsym\nu}$ are the Slater determinants where the set of orbitals $\bsym\mu$ and $\bsym\nu$ are occupied by the particles in $A$ and $B$. 

The reduced density matrix of $A$ is block diagonal with each block characterized by $(N_A,N_B)$ values. Calculation of a specific block 
\begin{equation}\label{rhoNAdefinition}
\rho_{N_A}={\rm Tr}_B\mathcal{P}_{N_A N_B}|\psi^{\rm unproj}_{n/(np+1)}\rangle\langle \psi^{\rm unproj}_{n/(np+1)} |\mathcal{P}_{N_A N_B}
\end{equation}
involves projecting the state into a sector where first $N_A$ particles are inside $A$ subsystem and $N_B$ are in $B$. The projected state is
\begin{equation}
\mathcal{P}_{N_A N_B}\psi^{\rm unproj}_{n/(np+1)}(\bsym{R})= \sum_{\vec{\bsym\mu},\vec{\bsym\nu}}s\rbkt{\vec{\bsym{\mu}};\vec{\bsym{\nu}}}
\xi_{\vec{\bsym\mu}}^{A}(\bsym{R}_A) \xi_{\vec{\bsym\nu}}^{B}(\bsym{R}_B)\label{NAprojection}
\end{equation}

Using Eq.\eqref{EWFexpansion}, Eq.\eqref{NAprojection} and Eq.\eqref{rhoNAdefinition} one can write the reduced density matrix for subsystem $A$ with $N_A$ as
\begin{equation} \label{rhoAMC}
	\rho_{N_A} = \sum_{\substack{ \vec{\bsym\mu}, \vec{\bsym\nu}\\  \vec{\bsym\mu}', \vec{\bsym\nu}'}} s(\vec{\bsym\mu};\vec{\bsym\nu}) s(\vec{\bsym\mu}';\vec{\bsym\nu}') \ket{\xi^{A}_{\vec{\bsym\mu}}} \iprod{\xi^{B}_{\vec{\bsym\nu}'}}{\xi^{B}_{\vec{\bsym\nu}}} \bra{\xi^{A}_{\vec{\bsym\mu}'}}
\end{equation}
It can be shown that the non-zero eigenvalues of $\rho_{N_A}$ is same as that of a matrix $M$ given by
\begin{equation} \label{MatrixM}
M_{\vec{\bsym\mu}, \vec{\bsym\nu}}= \sum_{\vec{\bsym\alpha},\vec{\bsym\beta}} s(\vec{\bsym\alpha};\vec{\bsym\beta}) s(\vec{\bsym\mu};\vec{\bsym\nu}) \iprod{\xi^{A}_{\vec{\bsym\alpha}}}{\xi^{A}_{\vec{\bsym\mu}}} \iprod{\xi^{B}_{\vec{\bsym\beta}}}{\xi^{B}_{\vec{\bsym\nu}}} 
\end{equation}
 Note that the overlaps of EWFs appearing here are computed inside their respective subsystems $A$ and $B$. We can find eigenvalues of an $L_z^A$ sector of $\rho_{N_A} $ by using only those EWFs which have angular momentum equal to $L_z^A$, in Eq.\eqref{rhoAMC}. 
For every $N_A$, there is a smallest possible relative momentum $L_{z0}^A$ for the $N_A$ particles inside $A$. In the remaining text, $L_z^A$ will represent the angular momentum of an EWF $\xi^A$ relative to $L_{z0}^A$.


For small systems, these overlap matrices can be computed exactly for both unprojected and projected CF states. For large systems, Monte Carlo methods are used to compute RSES for unprojected states.

In following two sections, we describe RSES computation for the cases of exact projection, and approximate projection using Monte Carlo methods, respectively.

\subsection{RSES for Exactly Projected CF states} \label{exactProj}

The methods described in Sec.~\ref{RSESAlgo} can be used to calculate the RSES of the projected CF states as well:
\begin{equation}\label{EWFexpansionProj}
\psi^{\rm proj}_{n/(np+1)}(\bsym{R})= {\rm anti}\sum_{\vec{\bsym\mu},\vec{\bsym\nu}}s\rbkt{\vec{\bsym{\mu}};\vec{\bsym{\nu}}}
[\mathcal{P}\xi_{\vec{\bsym\mu}}^{A}(\bsym{R}_A)] [\mathcal{P}\xi_{\vec{\bsym\nu}}^{B}(\bsym{R}_B)]
\end{equation}
which imply that the only change needed is that EWFs in Eq. \ref{EWFexpansion} should now be projected to the LLL:
	\begin{equation}
			 \proj\; {\mathcal{M}}_{\bsym{\mu^*}}(\bsym{R}) \sbkt{ {\mathcal{M}}_{\bsym{\mu}}(\bsym{R})}^{p}
	\end{equation}
Note that the results so far are exact. The exact RSES can be calculated if the LLL projection can be implemented exactly and if the matrix $M$ (Eq. \ref{MatrixM}) is exactly computed.

In the above expression, the ordered set $\bsym{\mu}^*$ may contain orbitals from higher LLs whereas $\bsym{\mu}$ contains orbitals in the LLL only. Exact LLL projection is implemented by elevating the Slater determinant $\mathcal{M}_{\bsym{\mu}^*}$ to an operator $\widehat{\mathcal{M}}_{\bsym{\mu}^*}$ which acts on $[\mathcal{M}_{\bsym{\mu}}]^p$. The operator $\widehat{\mathcal{M}}_{\bsym{\mu}^*}$ is constructed by replacing each single particle orbitals $\phi_{\mu^*}$ 
in the determinant 
$\mathcal{M}_{\bsym{\mu}}^*$ 
with an operator $\hat\phi_{\mu^*}$\cite{Jain07} that acts on LLL single particle orbitals as defined below.
In the spherical geometry, where the single particle orbitals are $Y_{Qnm}$ (Eq.\eqref{yqlm}), the operator that replaces this, acts on a LLL state $Y_{Q'0m'}$ as 
	\begin{equation}\label{SphereProj}
	\widehat{Y}^{Q'}_{Qnm}=\mathcal{N}_{QQ'}::{Y}_{Qnm}\rbkt{\bar{u}\rightarrow \partial_u,\bar{v}\rightarrow \partial_v} ::
	\end{equation}
where $::Y::$ represents a normal ordering where all $\bar{u}$ and $\bar{v}$ are moved to extreme right before replacing them with the corresponding derivatives. The coefficient $\mathcal{N}_{QQ'}$ is given by $\frac{2(Q+Q')+1}{2(Q+Q')+n+1}$.

It is computationally difficult to evaluate the projected state for large enough systems to perform the comparisons that we intend to do. Hence, we will consider the simplest non-trivial case namely that of $n=2, p=1$ which corresponds to a bosonic Jain state at filling fraction $2/3$. 
The EWFs (Eq.\eqref{EWFexpansion}) for this state can be written as 
	\begin{align} \label{EWFexpansionProj}
	\xi_{\vec{\bsym\mu}}^{A}(\bsym{R}_A)=& \widehat{\mathcal{M}}^{A}_{\bsym\mu^*} (\bsym{R}_A) \mathcal{M}^{A}_{\bsym\mu_1}(\bsym{R}_A)   \nonumber \\
	\xi_{\vec{\bsym\nu}}^{B}(\bsym{R}_B)=& \widehat{\mathcal{M}}^{B}_{\bsym\nu^*}(\bsym{R}_B)  \mathcal{M}^{B}_{\bsym\nu_1}(\bsym{R}_B)
	\end{align}
where the ordered sets $\bsym\mu^*$  and $\bsym\mu_1$ are $N_A$-sized subsets of orbitals in $\hat\Phi_2$ and $\Phi_1$ respectively. Similarly, $\bsym\nu^*$ and $\bsym\nu_1$ are $N_B$-sized complements of subsets $\bsym\mu^*$  and $\bsym\mu_1$ respectively.
The action of the operators $\hat{\phi}_{\mu^*_i}(\bsym{r}_i) $ a LLL state  ${\phi}_{\mu_{1,i}}(\bsym{r}_i)$, shown in Eq.\eqref{SphereProj}, can be written as 
\begin{gather}\label{coeffs}
\hat{\phi}_{\mu^*_i}{\phi}_{\mu_{1,i}}\equiv \hat{Y}^{Q'}_{Q{n_i}{m_i}}Y_{Q'0 m^{'}_i} =\nonumber \\= F(Q,Q',n_i,m_i,m^{'}_i) Y_{Q+Q',0,m_i+m^{'}_i}\equiv F(\mu^*_i, \mu_{1,i}){\phi}_{\gamma_i}
\end{gather}
where the coefficient $F(\mu^*_i, \mu_{1,i})$ for spherical geometry is given in the Appendix.
The projected  EWFs can be conveniently expanded in terms of  symmetric many particle states given by
\begin{widetext}
		\begin{align} \label{EWFsymm}
		\widehat{\mathcal{M}}^{A}_{\bsym\alpha} (\bsym{R}_A) \mathcal{M}^{A}_{\bsym\beta}(\bsym{R}_A) =&\,\frac{1}{N_A!}\sum_{P,Q\in \mathcal{S}_{N_A} } (-1)^{PQ} \prod_{i=1}^{N_A}\hat{\phi}_{\alpha_{P(i)}}(\bsym{r}_i) {\phi}_{\beta_{Q(i)}}(\bsym{r}_i) \nonumber \\
		=&\, \frac{1}{N_A!}\sum_{P\in \mathcal{S}_{N_A} } (-1)^{P}\, \sbkt{\prod_{i=1}^{N_A} F(\alpha(P)_{i}, \beta_i)} \text{sym}\rbkt{ \phi_{\gamma(P)_1}(\bsym{r}_1)\dots\phi_{\gamma(P)_{N_A}}(\bsym{r}_{N_A})}
	\end{align} 
\end{widetext}
where $\phi_{\bsym{\gamma}(P)_i}$ represents the state obtained from $\hat{\phi}_{\alpha_{P(i)}}\phi_{\beta_i}$ as shown in Eq.\eqref{coeffs}.
The symmetrization is defined as 
\begin{multline}\label{symmDef}
\text{sym}\rbkt{ \phi_{\lambda_1}(\bsym{r}_1)  \dots\phi_{\lambda_{N_A}}(\bsym{r}_{N_A})}=\sum_{P\in S_{N_A}} \prod_i \phi_{\gamma_i}(\bsym{r}_{P(i)})
\end{multline} 
We can expand all EWFs as linear combinations of orthogonal symmetrized basis states using Eq.\eqref{EWFsymm}. Using these, we can exactly compute matrix $M$ (Eq.\eqref{MatrixM}), without using Monte Carlo methods as we obtain the EWFs as an expansion in orthogonal states.
Eigenvalues of $M$ give the numerically exact RSES for the LLL projected $\psi_{2/3}$.

Exact projection can be performed on EWFs with $N_{A,B}\sim 12$ allowing calculation of exact RSES of the projected $2/3$ state in systems of sizes upto $N=N_A+N_B\sim 24$ particles.
We will now discuss possible approximations to the LLL projection of the EWFs which can be used instead of the exact projection. We will first discuss the Jain Kamila projection for the $2/3$rd state to motivate the approximations used.


\subsection{Jain-Kamilla projection} \label{JKoverview}
The LLL projected $2/3$ state can be evaluated in an approximate way as $\frac{1}{\Phi_1}\mathcal{P}\psi_{\frac{2}{5}}$. The LLL projected Jain state at filling fraction $2/5$ can be evaluated, again in an approximate way, using Jain Kamilla (JK) projection.\cite{Jain97b} The latter is computationally efficient and has been found to work well in variational studies.\cite{Jain07}
The LLL projected EWFs can be approximated in a similar manner. We give a short summary of the JK projection for the $2/5$ state first.

The CF state at filling fraction $2/5$ is given by
\begin{equation} \label{exactProjCF}
\proj\, {\Phi}_2(\bsym{R})\,\sbkt{ \Phi_1  (\bsym{R})}^{2}=\hat{\Phi}_2 (\bsym{R}) [\Phi_1 (\bsym{R})]^2
\end{equation}
which can be evaluated exactly
by elevating the first factor to an operator as described Sec.\ref{exactProj} but evaluating this explicitly is computationally expensive.

In the discussion below, we will only assume that $\Phi_2$ is a Slater determinant where particles occupy at most two Landau levels but not necessarily fully. $\Phi_1$ corresponds to LLL fully occupied by $N$ particles. $\sbkt{ \Phi_1  (\bsym{R})}^{2}$ appearing in the wavefunction above can be rewritten as
\begin{gather} \label{expandJastrow}
\sbkt{ \Phi_1  (\bsym{R})}^{2}= c\prod^{N}_i J_i,
\end{gather}
where $J_i= \prod^{N}_{k\neq i} \rbkt{u_i v_k - u_k v_i}$ and $c$ is some constant. With this, the unprojected CF state can be rewritten as
\begin{align} 
\psi^{\rm unproj}_{2/5}(\bsym{R})=& {\Phi}_2(\bsym{R})\,\sbkt{ \Phi_1  (\bsym{R})}^{2}= c\text{Det}\sbkt{Y_i(\Omega_j)} {\prod_i J_i} , \nonumber \\
=& c\text{Det}\sbkt{Y_i(\Omega_j)J_j}. \label{unprojCF3} 
\end{align}
where $\{Y_i(\Omega_j)\}$ represents the single particle orbitals in the Slater determinant $\Phi_2$. The JK projection approximates the projection of this determinant as follows 
\begin{align} \label{JK1} 
&\proj  \text{Det}\sbkt{Y_i(\Omega_j)J_j}\approx \text{Det}\sbkt{\proj Y_i(\Omega_j)J_j}=\text{Det}\sbkt{ \hat{Y}_i(\Omega_j)J_j}
\end{align}
where $\hat{Y}_i$'s are the operators defined in Eq.\eqref{SphereProj}. This can be simplified further by defining $\tilde{Y}$ as 
\begin{equation} 
J_j \tilde{Y}_i=\hat{Y}_i(\Omega_j)J_j.\label{ytilde}
\end{equation}
For the case, where $\Phi_2$ contains at most two lowest LLs (sufficient for our discussion), $\tilde{Y}_i$ can be shown to have the same form as Eq.\eqref{SphereProj}, but with the derivatives $\partial_{u},\,\partial_{v}$ in $\hat{Y}_i(\Omega_j)$ replaced as follows:
\begin{gather}
\partial_{u_j }\rightarrow \sum_{k\neq j}\,\frac{v_k}{u_j v_k - v_j u_k},\nonumber \\
\partial_{v_j}\rightarrow \sum_{k\neq j}\,\frac{-u_k}{u_j v_k - v_j u_k}. \label{approxJK}
\end{gather}
In summary, putting Eq.~\ref{ytilde} and Eq. \ref{JK1} together, the projected state can be approximated as  
\begin{equation}
 \hat{\Phi}_2 \sbkt{ \Phi_1  (\bsym{R})}^{2} \approx \sbkt{ \Phi_1  (\bsym{R})}^{2}\text{Det}\sbkt{ \tilde{Y}_i(\Omega_j)}
\end{equation}

\subsection{Approximate projection for Monte Carlo method} \label{approxProj}

The JK approximation relied on the possibility of writing the Jastrow Slater determinant $\Phi_1$ as a product of $J_i$ (Eq. \ref{expandJastrow}). The Slater determinants $\mathcal{M}^A_{\mu_1}, \mathcal{M}^B_{\mu_1}$ appearing in the EWFs (Eq. \ref{EWFexpansionProj}) do not have the same form as $\Phi_1$. We discuss an approximation similar in spirit to the JK projection but which applies to the EWFs. 

The EWFs occuring in the case of the RSES calculations of the projected $2/3$rd state (Eq.\eqref{EWFexpansionProj}) can be written as
\begin{equation}\label{eq33}
\xi^A=\mathcal{P} {\mathcal{M}}_{\bsym\alpha}^A   \mathcal{M}_{\bsym\beta}^A=\widehat{\mathcal{M}}_{\bsym\alpha}^A  \mathcal{M}_{\bsym\beta}^A.
\end{equation}
The state $\mathcal{M}_{\bsym\beta}^A$ is a Slater determinant of LLL orbitals:
\begin{equation}
\mathcal{M}_{\bsym\beta}^A = c\times {\rm anti}[\prod_{i=1}^{N_A}(u_i^{q+m_i} v_i^{q-m_i})]
\end{equation}
where $c$ is a constant, $q=(N-1)/2$, $m_i$ are the angular momenta of the single particle orbitals occupied inside $\mathcal{M}^A_{\bsym{\beta}}$. Factorizing out $\prod_i u_i^{2q}$, defining $z=u/v$, and using the definition of Schur polynomial we get 
\begin{equation}
\mathcal{M}_{\bsym\beta}^A = c  \times \prod_{i<j=1}^{N_A}(z_i-z_j) \times s(\{z_i\})\times \prod_{i=1}^{N_A} u_i^{2q}
\end{equation}
where $s$ is some Schur polynomial determined by $\bsym{\beta}$ and which has a small degree $\Delta\sim L_z^A$ relative to $N_A$.
Rewriting $z$ in terms of $u,v$ we get 
\begin{equation}
\mathcal{M}_{\bsym\beta}^A = \Phi_1^A f_{\bsym{\beta}}^A \prod_{i=1}^{N_A} u_i^{N-N_A-\Delta}
\end{equation}
where
\begin{equation}
\Phi_1^A=\prod_{i=1}^{N_A} \prod_{j=i+1}^{N_A} (u_i v_j - u_j v_i).
\end{equation}
The function $f$ is a polynomial in $u,v$ of degree $\Delta$. 

We can approximate the projected state, just as we did in the case of JK projection for $2/3$ state:
\begin{equation}
\xi^A \approx \frac{1}{\Phi^A_1}\widehat{\mathcal{M}}^A_{\bsym\alpha}  [\Phi^A_1]^2 f^A_{\bsym\beta} \prod_{i=1}^{N_A}u_i^{N-N_A-\Delta}\label{op1}
\end{equation}
Since $f$ has a small degree relative to other factors, we can ignore the action of the derivatives inside $\mathcal{M}^A_{\bsym{\alpha}}$ on $f$. 

Decomposing $[\phi_1^A]^2$ just as in the case of JK projection in terms of products of $J_i=\prod_{k=1,k\neq i}^{N_A}(u_iv_k-u_kv_i)$, 
we can make approximations similar to that in  Eq.\eqref{JK1}. The result is that we can replace the derivatives in $\widehat{\mathcal{M}}^A_{\bsym{\alpha}}$ with the following
\begin{align}
&\partial_{u_j }\rightarrow \frac{N-N_A}{u_j} +\sum_{k\neq j}\,\frac{v_k}{u_j v_k - v_j u_k},\nonumber \\
&\partial_{v_j}\rightarrow  \sum_{k\neq j}\,\frac{-u_k}{u_j v_k - v_j u_k}.\label{JKplusCorrection1}
\end{align}
which is a modification of the prescription given in Eq. \ref{approxJK}.

Similar approximations when made on $\xi^B$ result in the following prescription
\begin{align}
&\partial_{u_j}\rightarrow \sum_{k\neq j}\,\frac{v_k}{u_j v_k - v_j u_k},\nonumber \\
&\partial_{v_j}\rightarrow  \frac{N-N_B}{v_j} + \sum_{k\neq j}\,\frac{-u_k}{u_j v_k - v_j u_k}.\label{JKplusCorrection2}
\end{align}

In the above expressions, we have approximated $N-\Delta\approx N$. 
The additional factors appearing in the above prescription relative to the JK case (Eq. \ref{approxJK}) are associated with the fact that $\xi^A$ and $\xi^B$ have particles occupying only one hemisphere. The large multiplicative factors of $\prod_i u_i$ and $\prod_{i} v_i$ in $\xi^A$ and $\xi^B$ ensure that the net wavefunction is still in the LLL Hilbert space, in spite of the reciprocal terms $1/u_i$ and $1/v_i$.

The sequence of approximations made here is in spirit similar to the JK projection which do produce reliable results in variational studies in the context of energetics. Nevertheless, the approximations here go beyond JK projection and are being used in the context of RSES. Approximate projections can generate spurious effects such as destroy linear dependencies of trial functions,\cite{Rodriguez12b} which could in general change the counting of RSES. Therefore these calculations based on such approximations need quantitative comparisons with exact answers, which we do in the following section.

We consider other related, plausible approximations as well as results of the RSES calculated using them in the Appendix sections. These produce results that show less agreement with the exact RSES.
One of these is methods is to define the projected EWFs by simply replacing the derivatives using Eq. \ref{approxJK}. Note that when this is plugged into the right handsize of EWF expansion (Eq. \ref{EWFexpansionProj}), this is not yield the JK projected CF state. 

\begin{figure}[h!]	\includegraphics[width=0.9\columnwidth]{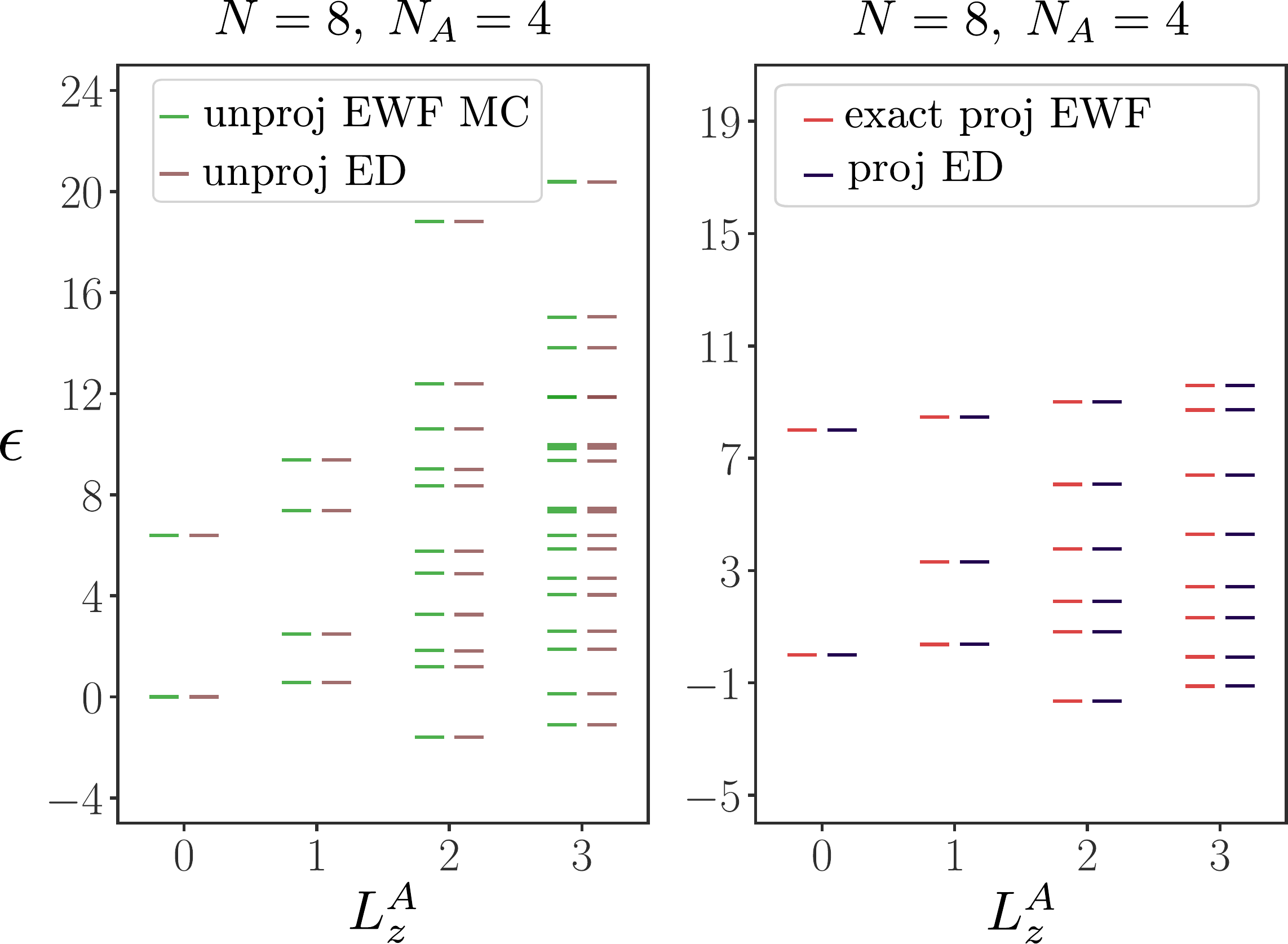}
	\caption{ 
		RSES for the unprojected (left panel) and projected (right panel) $2/5$ CF state, for $N=8$ and $N_A=4$. The left panel compares the RSES computed using methods in Sec.~\ref{RSESusingED} which uses ED ground state of TK Hamiltonian (brown) and the Monte Carlo from Sec.~\ref{RSESunproj} (green),  in given $L_z^A$-sectors.  For small $L_z^A$ values, both methods produce identical RSES (modulo an overall shift). In the right panel, we compare the RSES of the exact projected $2/5$ CF state computed using method in Sec.~\ref{exactProj} (red) with that of projected ground state of TK Hamiltonain, which uses method given in Sec.~\ref{RSESusingED} (yellow) for $N=8$ and $N_A=4$, in given $L_z^A$-sectors. Both methods produce identical RSES. All RSES are shifted so that the least entanglement energy at $L_z^A$ is equal to zero.
		\label{fig:Unproj25states}}
\end{figure}

\section{Numerical Results} \label{NumRes}
In this section we present the numerical results comparing the RSES computed using different methods.
To help keep track of which methods are being compared, we list them here
\begin{enumerate}
\item For RSES of the unprojected state obtained by using the Slater determinant expansion of the ground state of the TK Hamiltonian (Sec. \ref{RSESusingED}), we use the name ``unproj ED" in the figures. 

\item For RSES of the projected state obtained by using the Slater determinant expansion of the LLL projected ground state of the TK Hamiltonian (Sec. \ref{RSESusingED}), we use the name ``proj ED" in the figures.

\item For RSES of the unprojected state obtained by using Monte Carlo integrations of the unprojected EWFs (Sec. \ref{RSESunproj}), we use the name ``unproj EWF MC " in the figures.

\item For RSES of the projected state obtained by using exact projection of the EWFs as described in Sec. \ref{exactProj}, we use the name ``exact proj EWF" in the figures.

\item For RSES of the projected state obtained by using approximate projection (Sec. \ref{approxProj}), we use the name ``approx proj EWF MC" in the figures. 

\end{enumerate}

\subsection{Methods using EWFs give exact results}
In this section, we demonstrate that methods using EWFs indeed produce exact results. This is expected as there were no approximations made in the scheme. 

Figure~\ref{fig:Unproj25states} shows the RSES for unprojected (left panel) and projected (right panel) $2/5$ CF state with $N=8$ and $N_A=4$.
The RSES is computed using two methods. We use the method described in Sec.~\ref{RSESusingED} to compute the exact RSES of the ground state of the TK Hamiltonian obtained from ED. The second method uses MC integrations as described in Sec.~\ref{RSESAlgo}. As shown in the left panel, both methods produce exactly identical RSES. Note that the Monte Carlo method introduces an overall vertical shift in the RSES. We fix the shift by aligning both RSES such that lowest eigenvalue in the $L_z^A=0$ sector is equal to zero. The MC method costs a tiny fraction of the computational time and can be extended to 100s of particles (even at this large system size, the bottleneck is not the computational time but the numerical precision of the wavefunction evaluations).

In the right panel of Fig.~\ref{fig:Unproj25states}, we present the RSES for the projected $2/5$ state. Exact results are obtained by using the method in Sec.~\ref{RSESusingED} on the LLL projection of the ground state of the TK Hamiltonian.
Secondly, we compute the RSES by exactly projecting the EWFs for the $2/5$ CF state, as described in Sec.~\ref{exactProj}. As shown in the Fig.\ref{fig:Unproj25states}, since projection of the EWFs can be performed exactly, the second method produces exact results. Since computation of exact projection is feasible only for small number of particles, the methods work for small systems only.

\begin{figure}[h!]
\includegraphics[width=0.9\columnwidth]{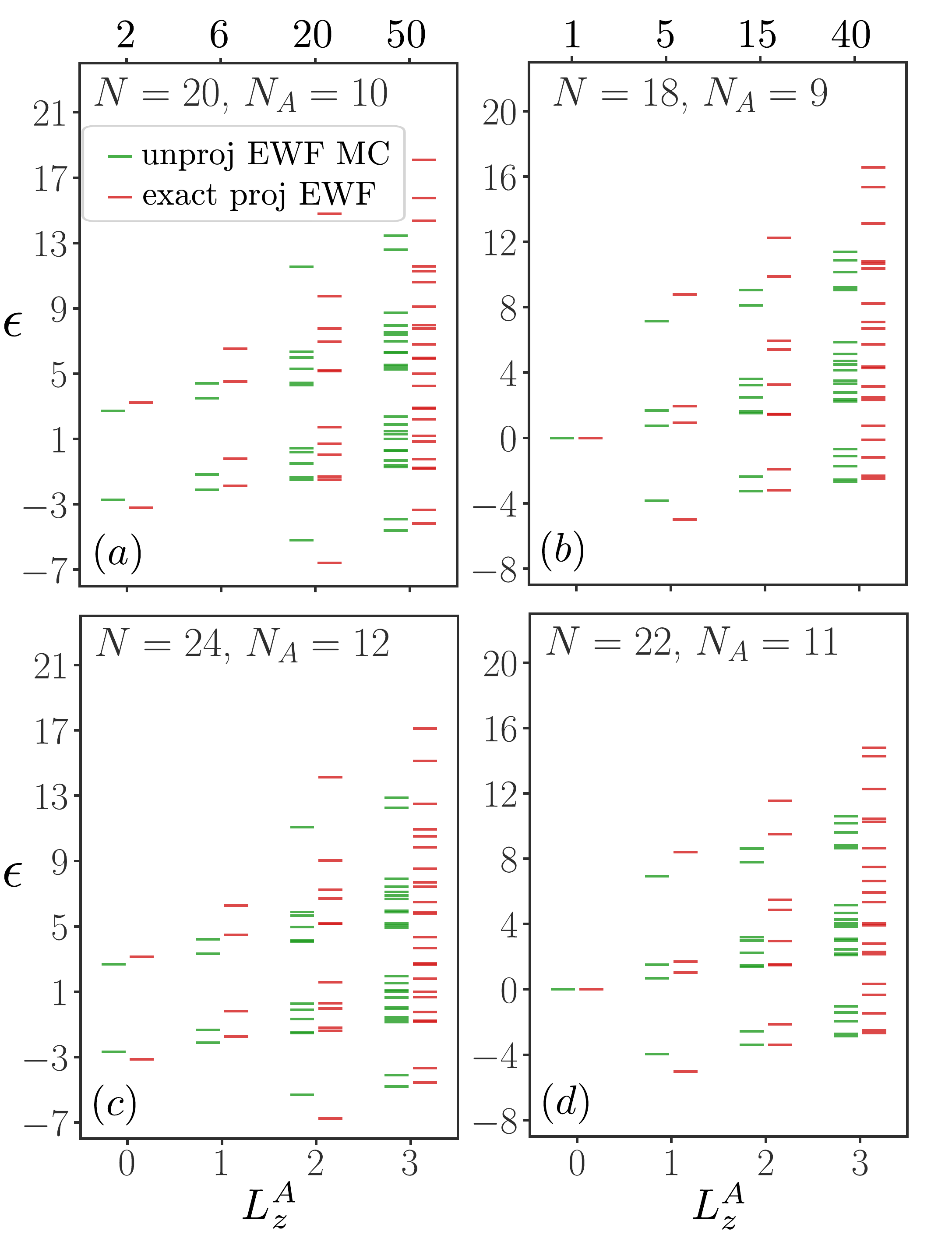}
\caption{RSES for the unprojected (green) and exactly projected (red) bosonic $2/3$ CF state. RSES for unprojected case is computed using Monte Carlo method from Sec.~\ref{RSESAlgo} whereas for exactly projected case, we use method presented in Sec.~\ref{exactProj}. RSES are shifted to make the mean entanglement energy equal to zero in $L_z^A=0$ sector. (a) and (c) show the RSES for even $N_A$ values, given by $N=20,\, N_A=10$ and $N=24,\, N_A=12$ respectively whereas (b) and (d) show the same for odd $N_A$ values,  given by $N=18,\, N_A=9$ and $N=22,\, N_A=11$ respectively. The numbers above the top axis of panel (a) and (b) represent the dimension of EWFs in each $L_Z^A$-sector for even and odd value of $N_A$, given the system is sufficiently large. Projected states have larger spread in the RSES for each $L_z^A$ sectors compared to RSES of unprojected state but the counting remains the same. 
\label{fig:RSEScomparison1}}
\end{figure}

\subsection{Effect of projection on RSES}
Here we compare the RSES of the projected and unprojected CF states in small systems, where we can obtain results exactly without any artifact from approximate projections. Calculation of exactly projected $2/5$ CF state is not feasible for larger systems. However, exact projection can be extended to larger systems $(N\sim 24)$ for the case of $2/3$ CF state.  We compare the RSES of the unprojected and (exact) projected CF state at $2/3$ in Fig.~\ref{fig:RSEScomparison1}. For the unprojected case we use a Monte Carlo method (which, as mentioned earlier, is practically exact) whereas exact RSES for the projected state is computed using the methods in Sec. \ref{exactProj}. RSES pattern is different for even and odd values of $N_A$ and we present the two largest system sizes for each. 

We see the RSES for the projected state shows the same counting as that of the unprojected state, though the RSES of the projected states show larger spread in the eigenvalues compared to that of the unprojected states. 
The numbers above the panels show the number of EWFs in each $L_z^A$ sector. The number of entanglement eigenvalues are smaller than that of the EWFs due to linear dependencies between them. The projection does not add further linear dependencies resulting in the same counting in both cases.

\begin{figure}[h!]
	\includegraphics[width=0.9\columnwidth]{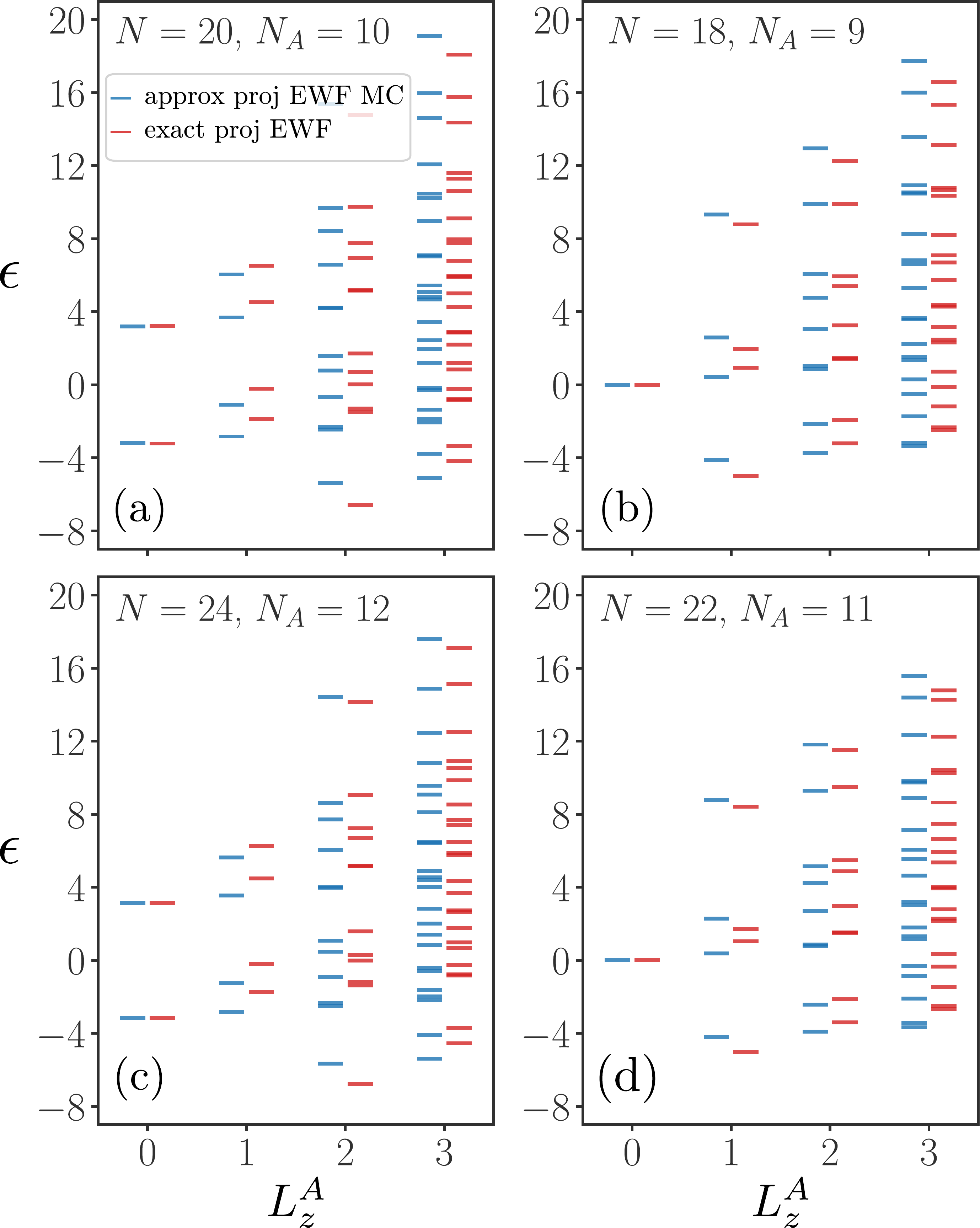}
	\caption{ RSES for exactly projected (red) and approximately projected (blue) bosonic $2/3$ CF state. RSES for the exactly projected state is compute using method in Sec.~\ref{exactProj} and method in Sec.~\ref{approxProj} is used for approximately projected state.   Since even and odd $N_A$ have different RSES pattern, we compare them separately. RSES in all $L_z^A$-sectors are shifted such that the mean entanglement energy at $L_z^A=0$ is equal to zero. Panel $(a)$ and $(c)$ show RSES comparison for system sizes $N=20,\,N_A=10$ and $N=24,\,N_A=12$ respectively. Panels (b) and (d) show similar comparisons for system sizes $N=18,\,N_A=9$ and $N=22,\,N_A=11$ respectively.
		\label{fig:RSEScomparison2}}
\end{figure} 

\subsection{RSES from approximate projection of EWFs}

In this section, we compare the exact RSES and the RSES obtained by approximating the LLL projection of the EWFs. As mentioned in the introduction, the comparison is not between RSES of exact projected CF state and RSES of the JK projected CF state - these two are likely to very similar to each other. The comparison is being made between the exact RSES and the RSES obtained after approximately projecting the EWFs.

Figure~\ref{fig:RSEScomparison2} shows the central result of our work comparing the RSES of exactly projected (Sec.~\ref{exactProj}) and approximately projected (Sec.~\ref{approxProj}) $2/3$ CF state for small systems. To make the comparison easier, all RSES are shifted vertically to make mean entanglement energy in $L_z^A=0$ sector equal to zero. 

Panels $(a)$-$(d)$ in Fig.~\ref{fig:RSEScomparison2} present a side-by-side comparison of RSES for approximately projected and exactly projected $2/3$ CF states for two largest systems where exact projection was feasible. Comparing the RSES for two different sizes, we see that the RSES for approximately projected state is close. Similar results were obtained even for smaller systems. Here too, the counting in different branches remain the same. It is noteworthy that the approximate projection does not destroy linear dependencies; this can be inferred from the identical counting in both cases.

A naive application of JK projection (using Eq. \ref{approxJK} instead of Eq.\eqref{JKplusCorrection1} and Eq.\eqref{JKplusCorrection2}) produces significant deviations from the exact RSES in small systems where comparisons can be performed. (See Fig. \ref{fig:RSEScomparisonA1} in Appednix).

\begin{figure}[h!]	\includegraphics[width=\columnwidth]{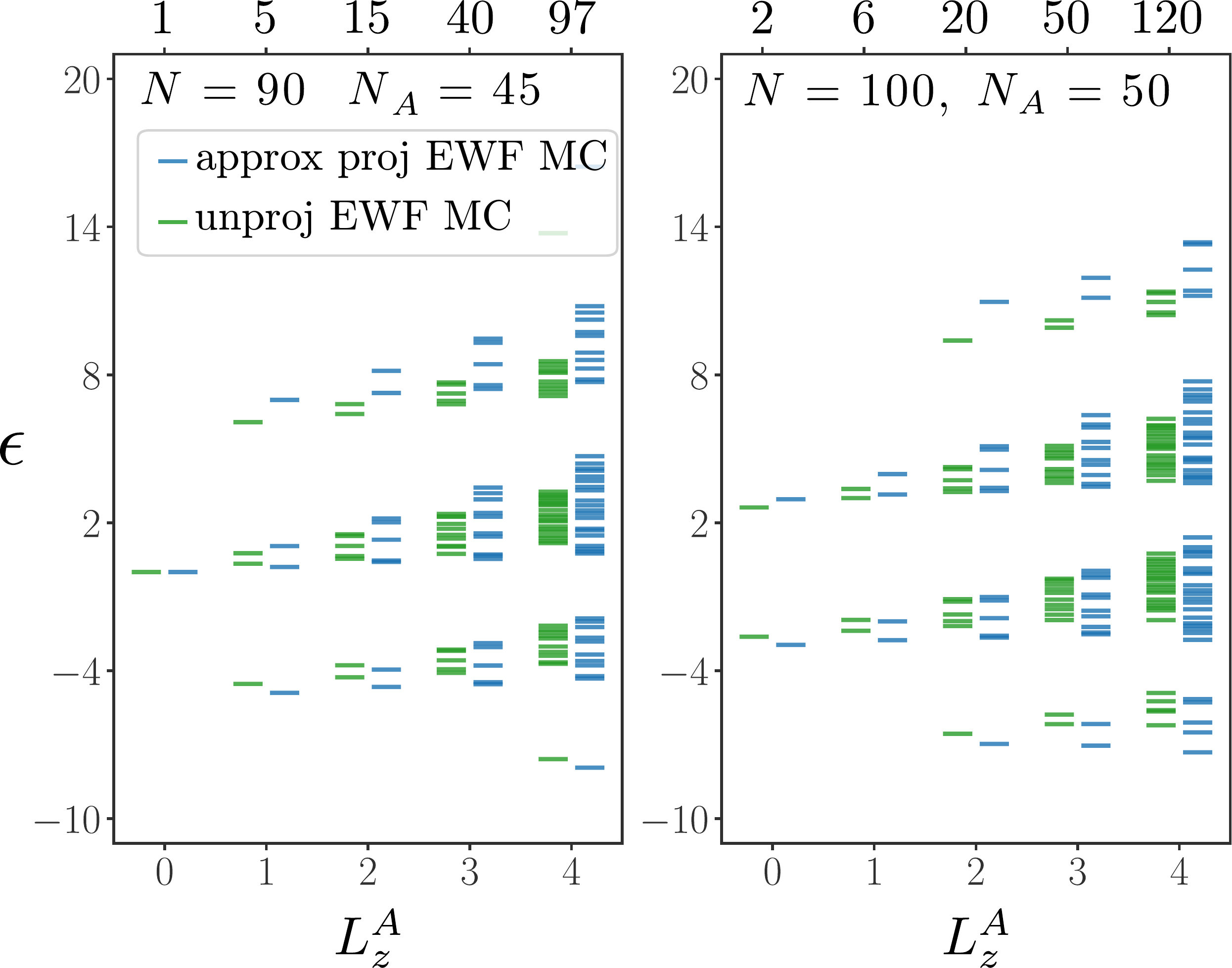}
	\caption{ 
	RSES for approximately projected (blue) and unprojected (green) bosonic $2/3$ CF for system sizes $N=90,\,N_A=45$ (left) and $N=100,\,N_A=50$ (right).  The RSES are shifted to make the mean entanglement energy in $L_z^A=0$ sector equal to zero. We see that, except for a larger spread, RSES for the projected state is qualitatively identical to that of the unprojected state for large systems. 
		\label{fig:projVSunproj}}
\end{figure}
	Finally, Fig.~\ref{fig:projVSunproj} shows the comparison between the unprojected and approximately projected $2/3$ CF state for two large system sizes $N=90,\,N_A=45$ (left) and $N=100,\,N_A=50$ (right). 	
Total number of EWFs in angular momentum-sectors given by $L_z^A=(0,1,2,3,4)$ for systems with even and odd $N_A$ are $(2,6,20,50,120)$ and $(1,5,15,40,97)$ respectively. Although the RSES for the projected state has a larger spread as was observed in the case of smaller systems in Fig.~\ref{fig:RSEScomparison1}, the branch structure is identical to that of the unprojected state. Qualitatively, the RSES of QH state does not change after projection. 	

\section{Conclusion} \label{conclusion}
In this work, we carefully analyzed the Monte Carlo methods for RSES calculations. MC methods are useful to explore RSES of variational states in FQHE as they can be used to study much larger systems (upto 100 particles) than what is possible with any of the other methods ($\sim 10$ particles). However when studying LLL projected CF states MC method needs adhoc approximations. This paper attempts to address the question of the quality of these approximations. We present alternate methods with which exact RSES can be computed in small systems and we used them to benchmark the MC results.

The MC method proceeds by working with EWFs which are those states in individual subsystems which `entangle' with the states in the complementary subsystem. These EWFs are far fewer in number when compared to the full dimension of the individual subsystems. \cite{Rodriguez12b,Rodriguez12,Rodriguez13}
In case of Jain CF states and parton states,\cite{Jain89,Jain89b} EWFs can be explicitly written down due to the specific structure of these functions. MC method uses the fact that the density matrix can be compactly expressed in a basis of made of the EWFs. In particular the eigenvalues of the density matrix is determined by the overlap matrix of the EWFs.

Firstly, by comparing with the exact RSES of the unprojected Jain 2/5th state, we have shown that the MC method, using a fraction of a computational cost, produces practically exact results. Although this is expected, a quantitave verification is appealing. Comparison could be performed only in small systems, however we believe that the exact agreement extends to large systems as well.

Progress can be made in the case of the RSES of projected states provided that suitable approximation schemes can be devised. 	
The EWFs have a structure different from the Jain states. It is not obvious that the naive application of the JK projection which works well in the Jain states would work for the EWFs. We present an approximation scheme for the EWFs that is in the spirit of the JK projection. In quantitative comparisons with the exact RSES of finite systems, the approximate method we presented works better than the one obtained by a naive application of the JK projection (the results of the latter are presented in the appendix). 

The approximate projection allows MC methods to be used in systems as large as 100 particles. We find that the RSES of the projected $2/3$rd state is very close to that of the unprojected state. Independent of the quality of the approximations used for the projection, we find that the counting of the RSES states remain robust. These suggest that all these methods produce close to the correct projected state.
Improved estimates of the RSES neverthelss can help in quantitative analysis of the scaling properties of the RSES. 

\bibliographystyle{apsrev}
\bibliography{biblio_fqhe.bib}

\widetext
\pagebreak

\begin{center}
	\textbf{\large Appendix}
\end{center}

\setcounter{equation}{0}
\setcounter{figure}{0}
\setcounter{table}{0}
\setcounter{page}{1}
\setcounter{section}{0}
\makeatletter
\renewcommand{\thesection}{A\arabic{section}}
\renewcommand{\theequation}{A\arabic{equation}}
\renewcommand{\thefigure}{A\arabic{figure}}
\renewcommand{\bibnumfmt}[1]{[A#1]}
\renewcommand{\citenumfont}[1]{#1}

\section{Coefficients for exact projection of 2/3 state on sphere}
We saw in Sec.\ref{exactProj}, computing the exact projection of bosonic $2/3$ CF state requires calculation of coefficients  $F(\alpha, \beta)$ arising when performing LLL projection on product of single particle orbitals:
\begin{equation}\label{A1}
\proj{\phi}_{\alpha}(\bsym{r}){\phi}_{\beta}(\bsym{r})=F(\alpha, \beta){\phi}_{\gamma}(\bsym{r})
\end{equation}
Here ${\phi}_{\beta}$ and ${\phi}_{\gamma}$ are LLL orbitals whereas ${\phi}_{\alpha}$ might occupy any arbitrary LL. For spherical geometry, Landau orbitals are represented by monopole harmonics $Y_{Q,n,m}(u,v)$\cite{Wu76,Wu77,Haldane83}. On the sphere, LHS of Eq.\eqref{A1} can be written as
\begin{equation}\label{A2}
	\proj{Y}_{Q,n,m}(u,v){Y}_{Q',0,m'}(u,v) =\hat{Y}_{Q,n,m}(u,v){Y}_{Q',0,m'}(u,v)
\end{equation}
where the operator $\hat{Y}_{Q,n,m}(u,v)$ is defined in Eq.\eqref{SphereProj}. Explicit expansion of the above expression produces a LLL orbital upto a proportionality constant:
\begin{equation}\label{A3}
	\hat{Y}_{Q,n,m}(u,v){Y}_{Q',0,m'}(u,v)=F(Q,Q',n,m,m')Y_{Q+Q',0,m+m'}(u,v)
\end{equation}
where the coefficients  $F(\alpha,\beta)\equiv F(Q,Q',n,m,m')$ are defined as
\begin{equation}\label{A4}
F(Q,Q',n,m,m')=\begin{cases} 
      \frac{ N^{LLL}_{Q,m} N^{LLL}_{Q',m'}}{N^{LLL}_{Q+Q',m+m'}} & n=0 \\
      -\frac{(2(Q+Q')+1)!}{(2(Q+Q')+2)!}\frac{N_{Q,n,m}N^{LLL}_{Q',m'}}{N^{LLL}_{Q+Q',m+m'}} \rbkt{{2Q+1 \choose Q-m+2    }(Q'-m') - {2Q+1 \choose Q+1-m}(Q'+m')} & n=1 
   \end{cases}
\end{equation}

where normalization $N_{Q,n,m}$ is defined in Eq.\eqref{yqlm} and normalization for state in LLL, $N^{LLL}_{Q,m}$, is given as
\begin{equation}
N^{LLL}_{Q,m}=N_{Q,0,m}\times {2Q \choose Q+m}\end{equation}

\section{Other approximations for LLL projection}
\begin{figure}[h!]
	\includegraphics[width=0.5\columnwidth]{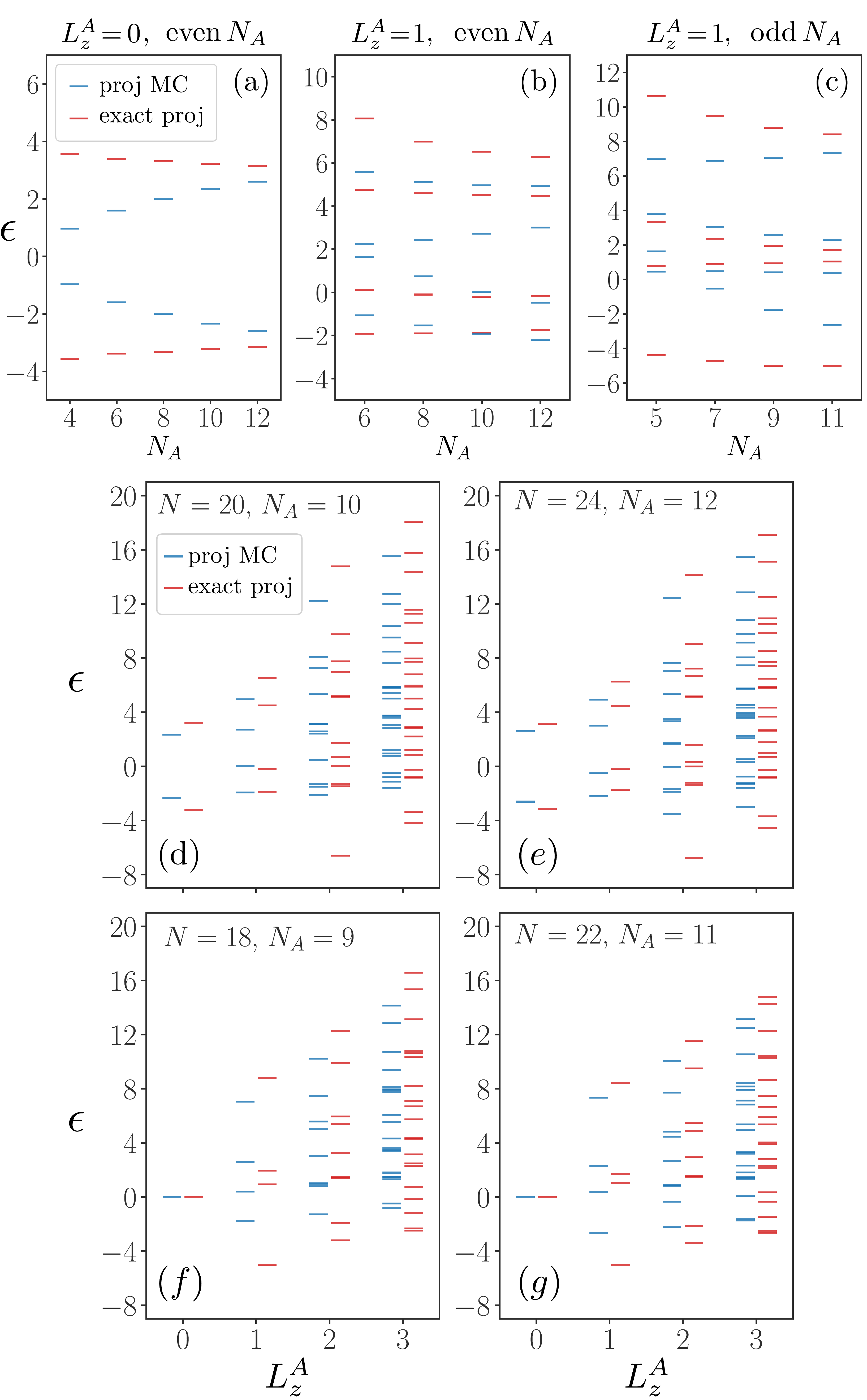}
	\caption{ RSES for exactly projected (red) and approximately projected (blue) bosonic $2/3$ CF state. RSES for the exactly projected state is compute using method in Sec.~\ref{exactProj} and method in Sec.~\ref{approxProj} is used for approximately projected state.   Since even and odd $N_A$ have different RSES pattern, we compare them separately. RSES in all $L_z^A$-sectors are shifted such that the mean entanglement energy at $L_z^A=0$ is equal to zero.  Panel $(a)$ shows the comparison of RSES for exactly and approximately projected states in $L_z^A=0$ sector. We see a clear convergence in RSES with $N_A$ here. Comparison in $L_z^A=0$ sector is trivial for odd $N_A$ values as there is only one entanglement energy. Panel $(b)$ and $(c)$ show RSES comparison in $L_z^A=1$ sector, for even and odd $N_A$ values respectively. Panel $(d)$ and $(e)$ show RSES comparison for system sizes $N=20,\,N_A=10$ and $N=24,\,N_A=12$ respectively. For same system size, RSES for approximately projected state has more spread compared to RSES of the exactly projected state. Panels (f) and (g) show similar comparisons for system sizes $N=18,\,N_A=9$ and $N=22,\,N_A=11$ respectively.
		\label{fig:RSEScomparisonA1}}
\end{figure}

In this section, we present the results comparing RSES of bosonic Jain $2/3$ state computed using two different approximations for the LLL projection. First, we use the JK method given in Sec.\ref{JKoverview}, where LLL projection is equivalent to following replacement of derivatives
\begin{gather}
\partial_{u_j }\rightarrow \sum_{k\neq j}\,\frac{v_k}{u_j v_k - v_j u_k},\nonumber \\
\partial_{v_j}\rightarrow \sum_{k\neq j}\,\frac{-u_k}{u_j v_k - v_j u_k}. \label{approxJK_App}
\end{gather}
in all operators $\widehat{Y}_{Q,n,m}$, in Eq.\eqref{SphereProj}. Fig.\ref{fig:RSEScomparisonA1} shows the comparison of RSES generated using this method with that of the exactly projected state. Both RSES are shifted to make the mean entanglement energy  in $L_z^A=0$ sector equal to zero. Panel $(a),\ (b)$ and $(c)$ show the trend of entanglement energies for the two methods with increasing system size, in given $L_z^{A}$-sectors.   We see, with increasing system size,  both RSES appear to converge towards each other. Panel $(d)-(g)$ show RSES comparison for fixed system size, each one for largest system sizes accessible for the exact projection. Compared to RSES (Fig.\ref{fig:RSEScomparison2}) of the approximation given in Sec.\ref{approxProj}, we see larger mismatch with RSES of exactly projected state, which worsens in larger $L_z^A$-sectors. It is also apparent that the mismatch decreases with increasing system size.

\begin{figure}[h!]
	\includegraphics[width=0.4\columnwidth]{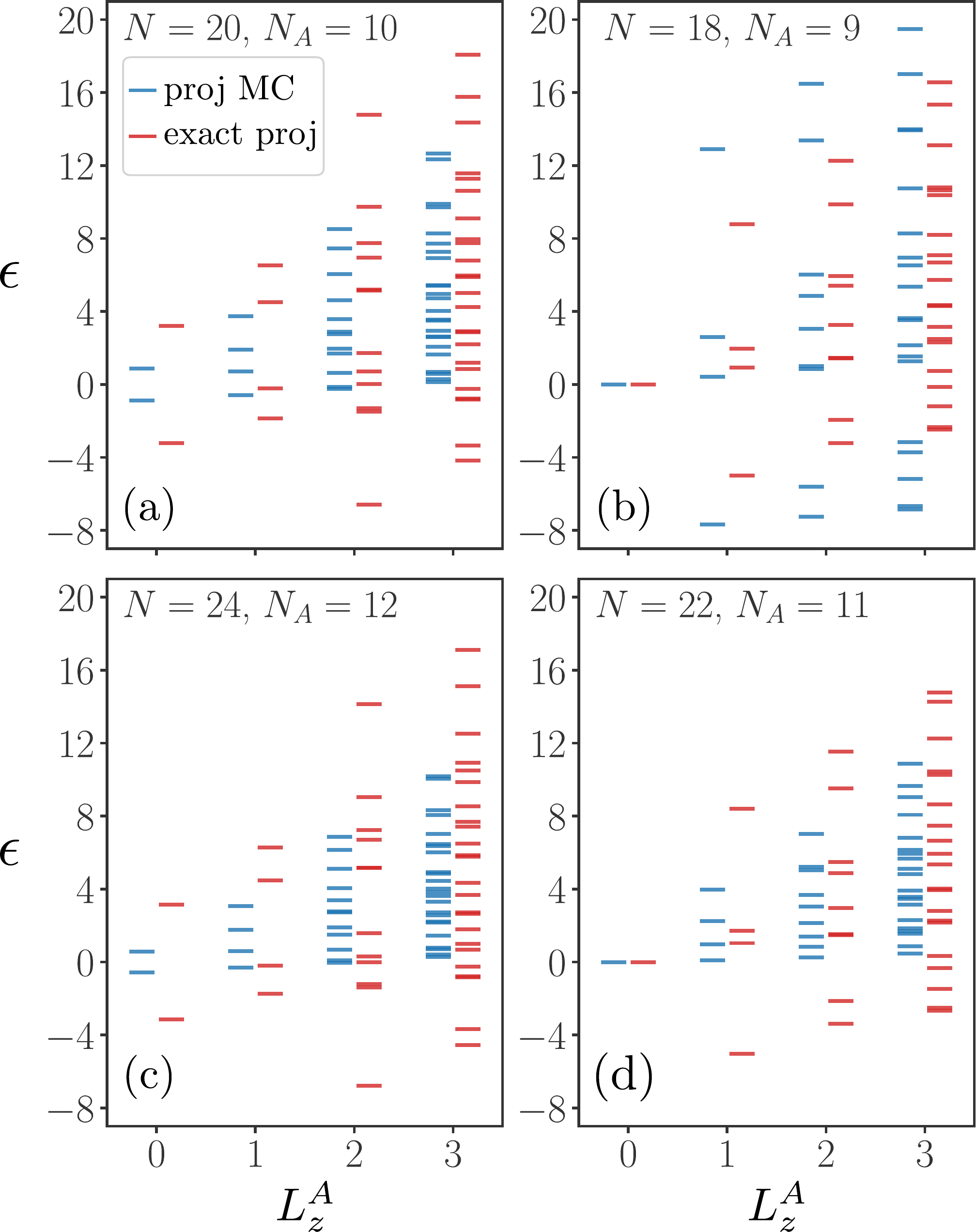}
	\caption{ RSES for exactly projected (red) and approximately projected (blue) bosonic $2/3$ CF state. RSES for the exactly projected state is compute using method in Sec.~\ref{exactProj} and method in Sec.~\ref{approxProj} is used for approximately projected state.  Panel $(a)$ and $(c)$ show RSES comparison for system sizes $N=20,\,N_A=10$ and $N=24,\,N_A=12$ respectively. For same system size, RSES for approximately projected state has no clear entanglement gaps and does not look similar to the RSES of the exactly projected state. Panels (b) and (d) show similar comparisons for system sizes $N=18,\,N_A=9$ and $N=22,\,N_A=11$ respectively.
		\label{fig:RSEScomparisonA2}}
\end{figure}

For the second case, we approximate the projection of EWFs as
\begin{equation}
\xi^A =\mathcal{P} {\mathcal{M}}_{\bsym\alpha}^A   \mathcal{M}_{\bsym\beta}^A=\widehat{\mathcal{M}}_{\bsym\alpha}^A  \mathcal{M}_{\bsym\beta}^A\approx \frac{1}{\mathcal{M}_{\bsym\beta}^A}\widehat{\mathcal{M}}^A_{\bsym\alpha}  [\mathcal{M}_{\bsym\beta}^A]^2\label{op2}
\end{equation}
This approximation leads to following replacement of derivatives, given by
\begin{align}
&\partial_{u_j }\rightarrow 2\frac{N-N_A}{u_j} +\sum_{k\neq j}\,\frac{v_k}{u_j v_k - v_j u_k},\nonumber \\
&\partial_{v_j}\rightarrow  \sum_{k\neq j}\,\frac{-u_k}{u_j v_k - v_j u_k}.
\end{align}
for $A$-subsection and for $\xi^B$, results in the following prescription
\begin{align}
&\partial_{u_j}\rightarrow \sum_{k\neq j}\,\frac{v_k}{u_j v_k - v_j u_k},\nonumber \\
&\partial_{v_j}\rightarrow  2\frac{N-N_B}{v_j} + \sum_{k\neq j}\,\frac{-u_k}{u_j v_k - v_j u_k}.
\end{align}
Fig.\ref{fig:RSEScomparisonA2} shows the similar comparison of resultant RSES with that of exactly projected Jain $2/3$ state, for different system sizes. We notice that the RSES is significantly different than that of exactly projected state, with no clear entanglement gaps  even in small $L_z^A$-sectors. 

\end{document}